\documentclass{main}
\usepackage{amsmath,amssymb}
\usepackage{siunitx}
\usepackage{graphicx}
\usepackage{booktabs}
\usepackage{microtype}
\usepackage{xspace}
\usepackage{hyperref}
\usepackage[capitalise]{cleveref}
\usepackage{textcomp}
\usepackage{fancyhdr}
\usepackage{lastpage}
\hypersetup{
  colorlinks=true,
  linkcolor=blue,
  citecolor=blue,
  urlcolor=blue
}

\def\be{\begin{equation}}
\def\ee{\end{equation}}
\def\bea{\begin{eqnarray}}
\def\eea{\end{eqnarray}}

\newcommand{\elel}{\ensuremath{e^+e^-}\xspace}

\newcommand{\tautau}{\ensuremath{\tau^+\tau^-}\xspace}
\newcommand{\gaga}{\ensuremath{\gamma\gamma}\xspace}
\newcommand{\pp}{\ensuremath{pp}\xspace}
\newcommand{\PbPb}{\ensuremath{\mathrm{Pb\text{--}Pb}}\xspace}
\newcommand{\snn}{\ensuremath{\sqrt{s_{\mathrm{NN}}}}\xspace}
\newcommand{\pT}{\ensuremath{p_{\mathrm{T}}}\xspace}

\newcommand{\GeV}{\si{GeV}\xspace}
\newcommand{\GeVc}{\si{GeV}/c\xspace}
\newcommand{\nbinv}{\ensuremath{\mathrm{nb}^{-1}}\xspace}
\newcommand{\order}{\ensuremath{\mathcal{O}}\xspace}
\newcommand{\ALICE}{ALICE\xspace}
\fancypagestyle{firstpagefooter}{%
  \fancyfoot[L]{  \textcopyright \footnotesize This work is an open access article under a Creative Commons Attribution 4.0 International License (https://creativecommons.org/licenses/by/4.0/).}
  \fancyfoot[C]{}  % **Red: This removes the page number from the first page footer**
   
}

\begin{document}

\thispagestyle{firstpagefooter}
\title{\Large Measurement of the tau anomalous magnetic moment using Ultra-peripheral collisions with the \ALICE\ detector in~Run~3 \PbPb\ data}

\author{\underline{R.~Lavi\v{c}ka}\footnote{Speaker, email: roman.lavicka@cern.ch} and P.~A.~B\"{u}hler for the ALICE Collaboration}

\address{
Stefan Meyer Institute for Subatomic Physics, Austrian Academy of Sciences\\
Wiesingerstraße 4, 1010 Wien, Austria
}

\maketitle\abstracts{
The anomalous magnetic moment of the tau lepton ($a_{\tau}$) is a sensitive probe for the search for deviations from the Standard Model predictions and thus for new physics. This study investigates the feasibility of measuring $a_{\tau}$ using ultra-peripheral collisions (UPCs) at the LHC, where photon-photon interactions ($\gamma\gamma \to \tau^+ \tau^-$) produce tau lepton pairs. We focus on events recorded by the ALICE detector during Run 3 Pb—Pb data-taking. Events are selected in the decay channel where one tau decays into an electron  and neutrinos, and the other decays into a charged pion, or three charged pions, and neutrinos. These samples are enhanced with decays into muons, which are inseparable in the ALICE detector. The clean environment of UPCs minimizes hadronic background, while the advanced particle identification capabilities of the ALICE Time Projection Chamber (TPC) and Time-of-Flight (TOF) systems allow for efficient separation of electrons, pions, and background particles. In this talk, prospects for measuring this process by ALICE in Run 3, which benefits from high statistics and improved systematics uncertainties, will be discussed. Results will provide tighter constraints on $a_{\tau}$, contributing to the broader effort to test the Standard Model's robustness and explore physics beyond it.
}

\footnotesize DOI: \url{https://doi.org/xx.yyyyy/nnnnnnnn}


\section{Introduction}
For a point-like Dirac fermion with spin $1/2$, the gyromagnetic factor is $g=2$. Quantum corrections introduced by Standard Model (SM) induce the \emph{anomalous magnetic moment},
\begin{equation}
  a \equiv \frac{g-2}{2}.
\end{equation}
In case of leptons, electron $a_e$~\cite{Hanneke:2010au} and muon $a_\mu$~\cite{Muong2:2023} are measured with extraordinary precision, the tau lepton remains poorly constrained due to its short lifetime and the consequent absence of spin-precession based measurements. Current limits are derived from high-energy production processes and leave orders of magnitude of parameter space unconstrained relative to the electron and muon~\cite{CMS:2024RPP}.

Given that sensitivity to new physics grows with the lepton mass squared, a precise determination of $a_\tau$ would strongly constrain TeV-scale scenarios beyond the SM. Ultraperipheral heavy-ion collisions (UPCs) at the LHC provide an intense flux of quasi-real photons with very small virtuality ($q^2\to 0$), enabling studies of $\gaga\to\tautau$ that are directly sensitive to the $\gamma\tau\tau$ vertex~\cite{Berger:1991}.

\section{Theoretical Background and Status}
The effective $\gamma\tau\tau$ vertex can be written as~\cite{Beresford:2020,Dyndal:2020}
\begin{equation}
\Gamma^\mu_{\gamma\tau\tau}(q) = -ie\left[ \gamma^\mu F_1(q^2) + \frac{i}{2m_\tau}\sigma^{\mu\nu}q_\nu F_2(q^2) + \frac{1}{2m_\tau}\gamma^5\sigma^{\mu\nu}q_\nu F_3(q^2)\right],
\end{equation}
with $F_2(0) = a_\tau$.
Constraints on $a_\tau$ have been derived from \elel\ colliders (e.g.\ DELPHI at LEP~\cite{DELPHI:2001}), LHC \pp\ data (CMS~\cite{CMS:2024RPP}), and heavy-ion UPCs (ATLAS~\cite{ATLAS:2022ryk} and CMS~\cite{CMS:2022arf}), probing complementary $\gaga$ center-of-mass energies $W_{\gaga}$: below \SI{10}{GeV} at LEP, $\order(10$--$50)$~\GeV\ in \PbPb, and up to several hundred \GeV\ in \pp. The overall best bounds to date come from CMS in \pp, while the \ALICE\ program targets the intermediate $W_{\gaga}$ region with abundant low-\pT\ tracks and excellent PID at midrapidity. While in Run 2 the limited luminosity and detector performance did not allow a meaningful measurement of $a_\tau$, the upgraded ALICE detector and larger datasets of Run 3 open the possibility to perform this measurement for the first time.

\section{Experimental Setup and Run 3 Performance}
\begin{figure}[b]
  \centering
  % Placeholder figure box (compiles without image files)
  % \fbox{\rule{0pt}{1.6in}\rule{0.9\linewidth}{0pt}}
  \includegraphics[width=0.48\textwidth]{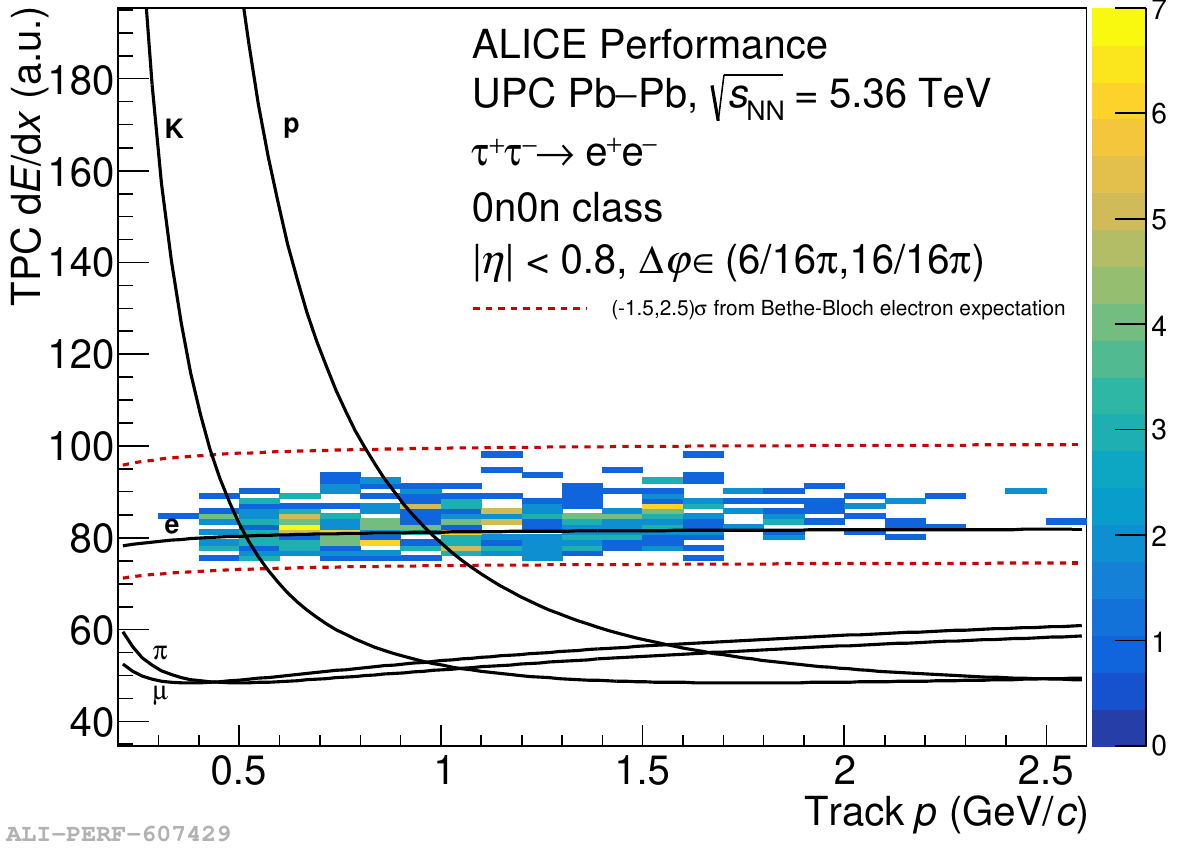}\includegraphics[width=0.48\textwidth]{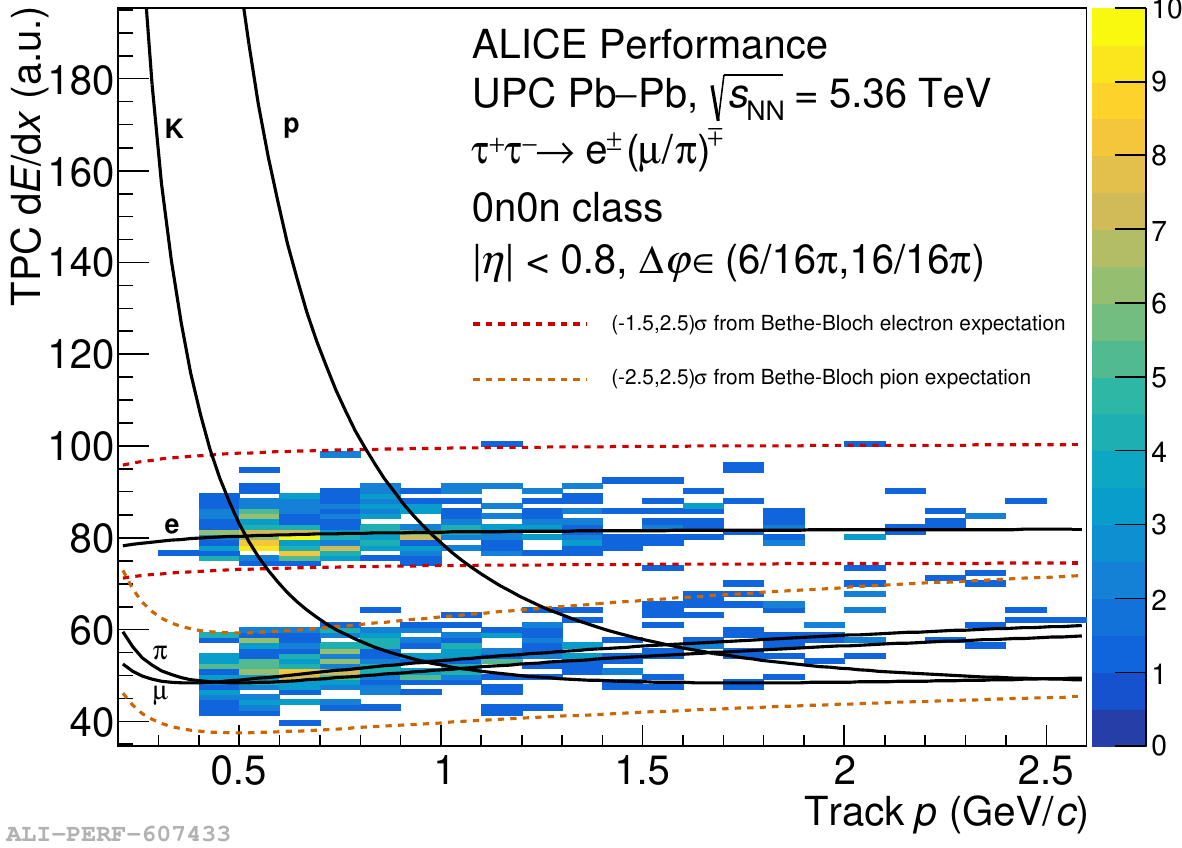}
  \caption{TPC $\mathrm{d}E/\mathrm{d}x$ vs.\ track momentum distributions showing electron--muon/pion separation and the absence of kaon/proton contamination in the selected samples. Black curves indicate Bethe-Bloch expectations for various particles.}
  \label{fig:pid}
\end{figure}

The \ALICE\ detector underwent major upgrades for Run~3~\cite{ALICE:JINST2024}. Subsystems relevant to this analysis include:
\begin{itemize}
  \item \textbf{DAQ:} the triggered readout of all relevant detectors was replaced with streaming readout;
  \item \textbf{ITS:} improved vertexing and low-\pT\ tracking;
  \item \textbf{TPC:} main tracking detector with excellent $\mathrm{d}E/\mathrm{d}x$ resolution for PID of low-momenta tracks;
  \item \textbf{TOF:} timing-based PID extending electron--kaon/proton separation up to $\pT\sim\SI{1.5}{\GeVc}$;
  \item \textbf{FIT:} quartz radiators and scintillators at forward rapidities for interaction time and multiplicity;
  \item \textbf{ZDC:} forward hadronic calorimeters to detect neutrons.
\end{itemize}
In the central barrel, electrons are cleanly separated from muons, pions, kaons and protons with TPC+TOF for $\pT<\SI{1.5}{\GeVc}$, see Figure~\ref{fig:pid}. Pions and muons are not separable via TPC/TOF alone, but this does not limit the mixed electron$+$pion/muon channel used below.

\section{Analysis Strategy}
Tau decays are categorized into 1-prong ($\sim80\%$) and 3-prong ($\sim20\%$) modes. \ALICE\ focuses on final states with an \emph{electron} and an oppositely charged track system:
\begin{itemize}
  \item \textbf{1-prong:} $e^\pm + e^\mp\ {\rm or}\ (\pi^\mp/\mu^\mp)$;
  \item \textbf{3-prong:} $e^\pm + 3\pi^\mp$ (with neutral pions unconstrained).
\end{itemize}
The dominant backgrounds are $\gaga\to\elel$, $\gaga\to\mu^+\mu^-$, $\gaga\to\pi^+\pi^-$ and continuum coherent photonuclear production of $\pi^+\pi^-$. The selection employs:
\begin{enumerate}
  \item \textbf{Exclusivity:} two-prong topology at midrapidity with minimal additional activity;
  \item \textbf{Acoplanarity:} back-to-back pairs veto to suppress continuum; a requirement on system $\pT$ (e.g.\ $>\SI{0.2}{\GeVc}$) is used to reduce \elel-contamination;
  \item \textbf{PID:} TPC $\mathrm{d}E/\mathrm{d}x$ and TOF to tag electrons and reject hadrons.
  \item \textbf{No neutrons:} veto on events with neutrons in forward direction to reject nuclear break-up events (in figures referred as 0n0n class).
\end{enumerate}
Residual background is estimated using like-sign and mixed-event techniques. For the $e+3\pi$ channel, an invariant-mass selection $m_{3\pi}<m_\tau$ further suppresses combinatorics.

\section{Simulation Studies}
Dedicated simulations of UPC \PbPb\ collisions at \snn$=\SI{5.02}{TeV}$ based on Upcgen generator~\cite{Burmasov:2021phy} corresponding to integrated luminosities of $2.7$--$13$~\nbinv\ predict $\sim 3.6\times10^4$ \tautau\ events within the central-barrel acceptance when requiring one electron and one oppositely charged track (pion/muon) at midrapidity. The relative yield fractions for the accessible channels are approximately $(e+e):(e+\pi/\mu):(e+3\pi)=1:5.6:1.7$.

Sensitivity to $a_\tau$ is driven by distortions of the lepton \pT\ spectra. We form ratios of the differential cross section across 10 \pT\ intervals and perform a $\chi^2$ combination assuming uncorrelated systematics. Depending on the achievable systematic uncertainty (1--5\%), the projected \ALICE\ limits improve upon the DELPHI constraint by factors of $2$--$8$ with the full Run~3+4 dataset.

\section{First Results from Run 3}
\begin{figure}
  \centering
  \includegraphics[width=0.48\textwidth]{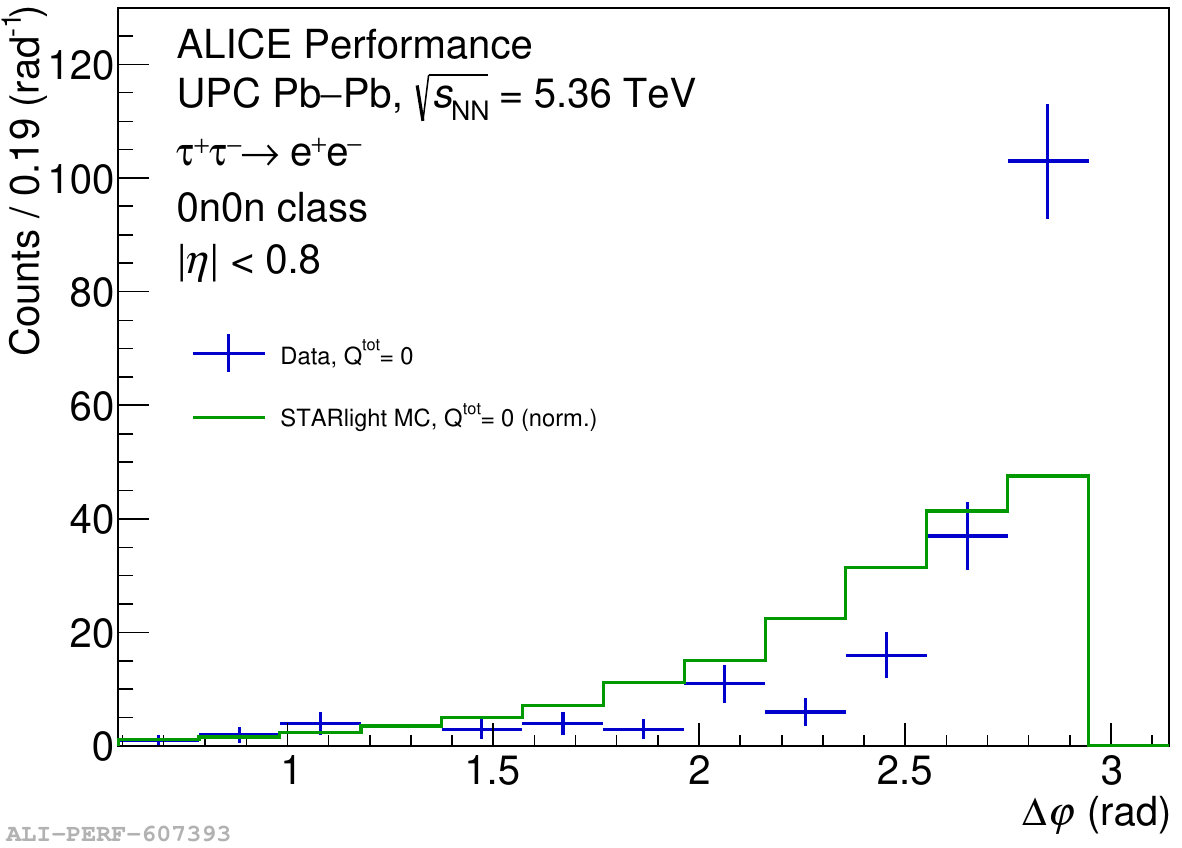}\includegraphics[width=0.48\textwidth]{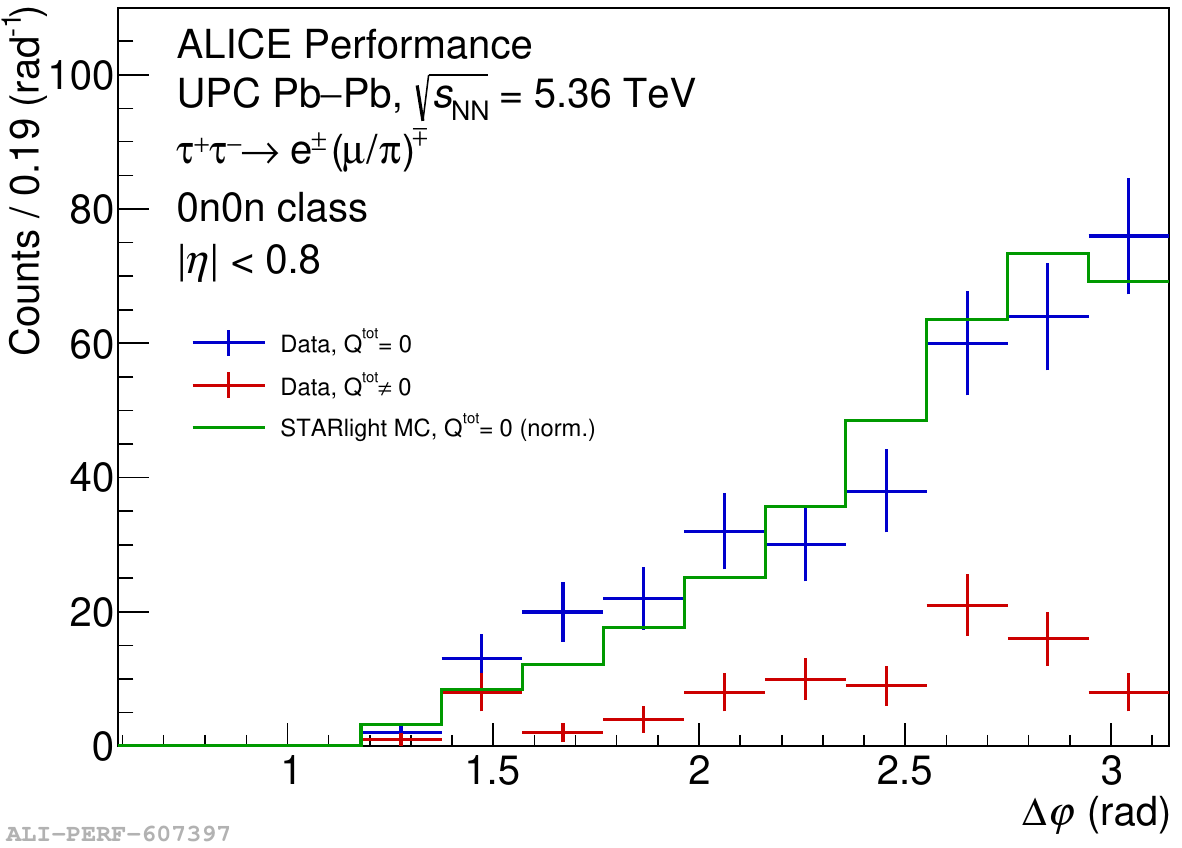}
  \caption{Acoplanarity distributions in data vs.\ STARlight MC for the \elel and $e+(\pi/\mu)$ channels after requiring system $\pT>\SI{0.2}{\GeVc}$.`}
  \label{fig:acoplanarity}
\end{figure}

\begin{figure}
  \centering
  \includegraphics[width=0.48\textwidth]{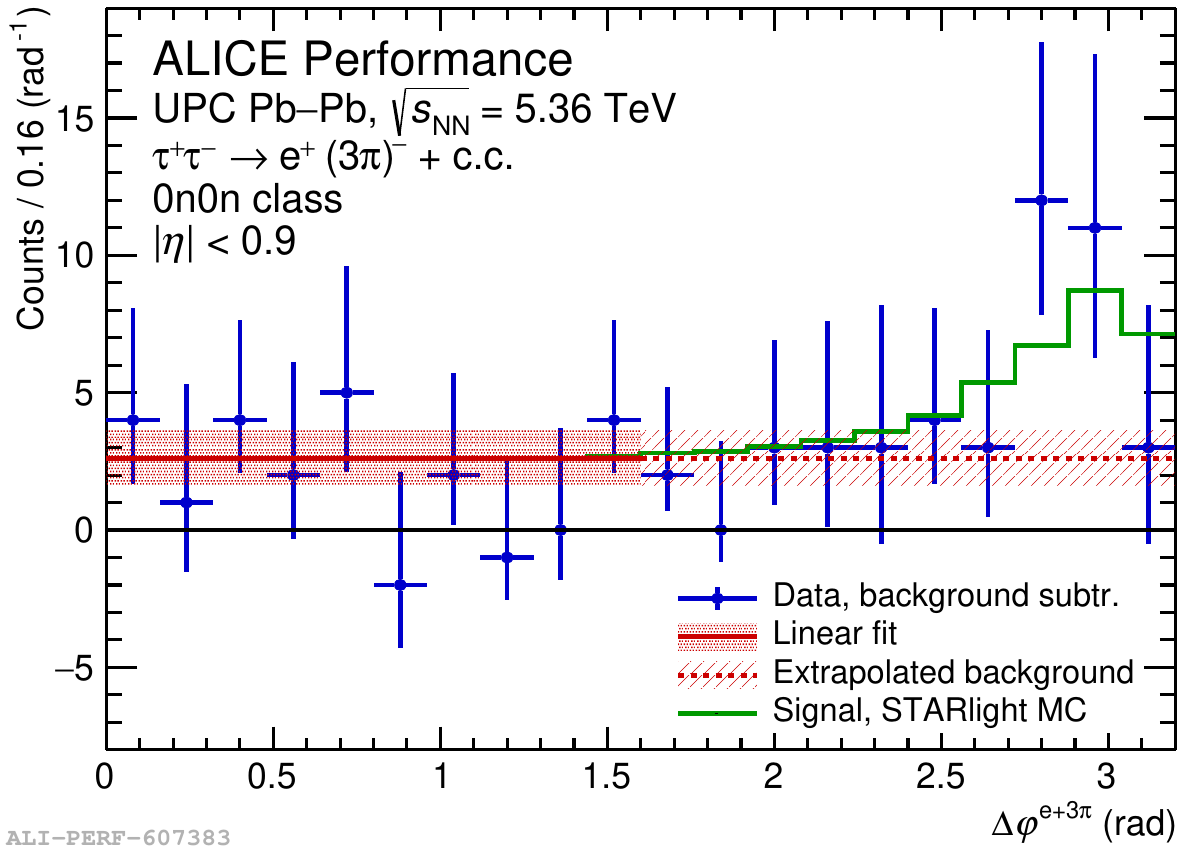}\includegraphics[width=0.48\textwidth]{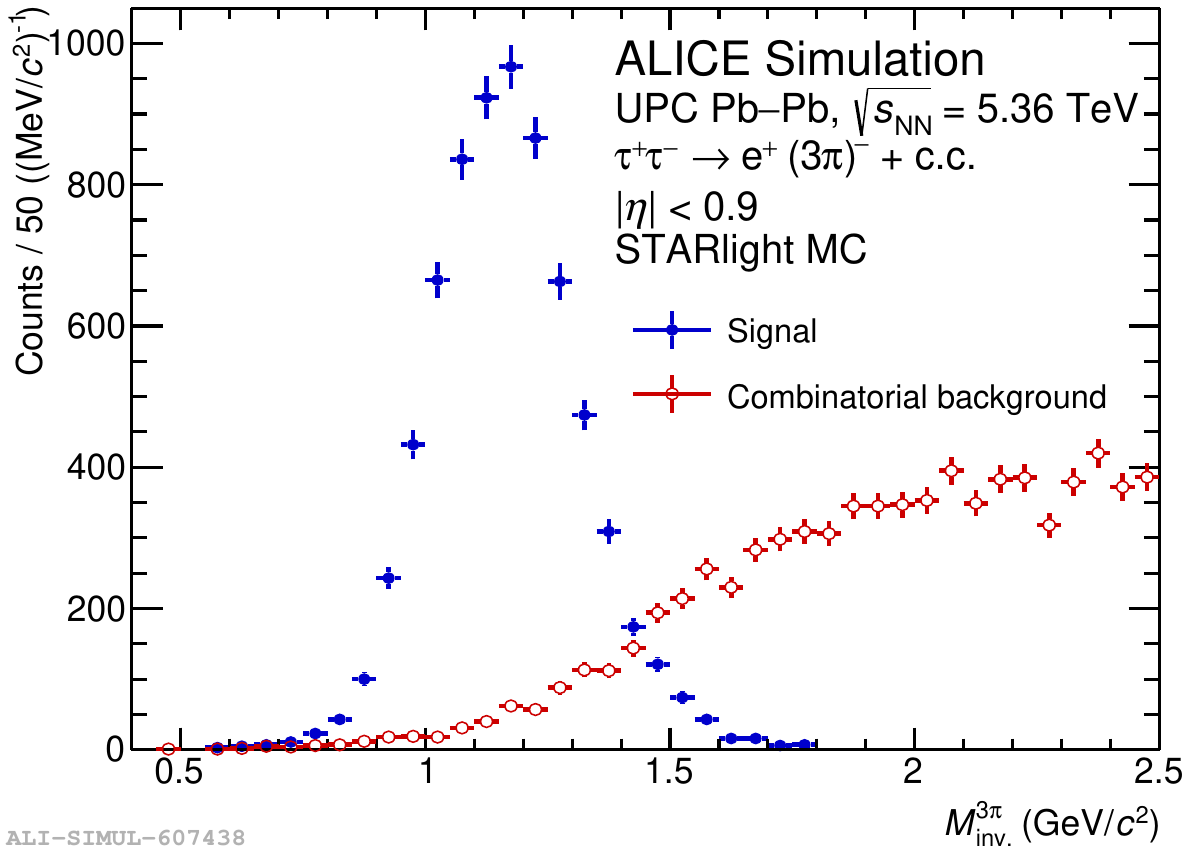}
  \caption{Acoplanarity (left) and invariant mass (right) distributions in data vs.\ STARlight MC for the \elel and $e+3\pi$ channel after requiring system $\pT>\SI{0.2}{\GeVc}$.`}
  \label{fig:fourprong}
\end{figure}

\begin{figure}
  \centering
  \includegraphics[width=0.48\textwidth]{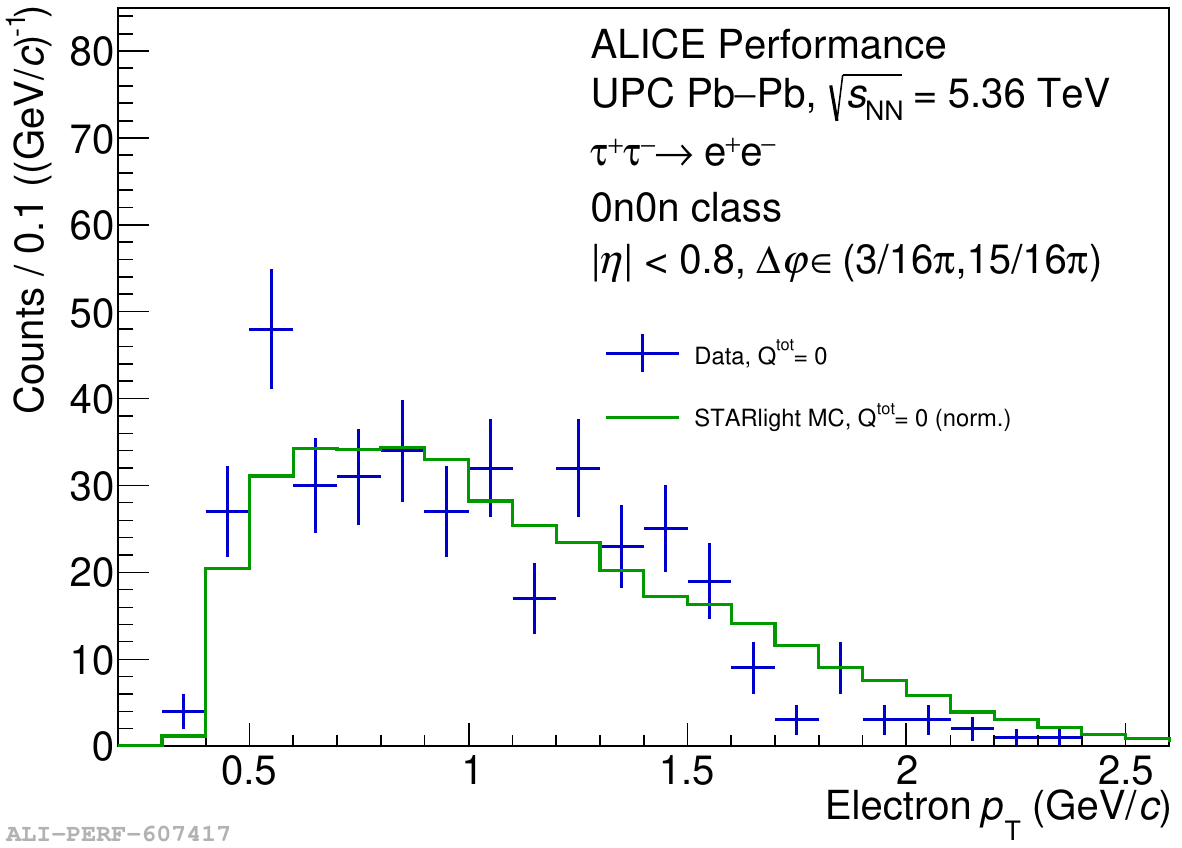}\includegraphics[width=0.48\textwidth]{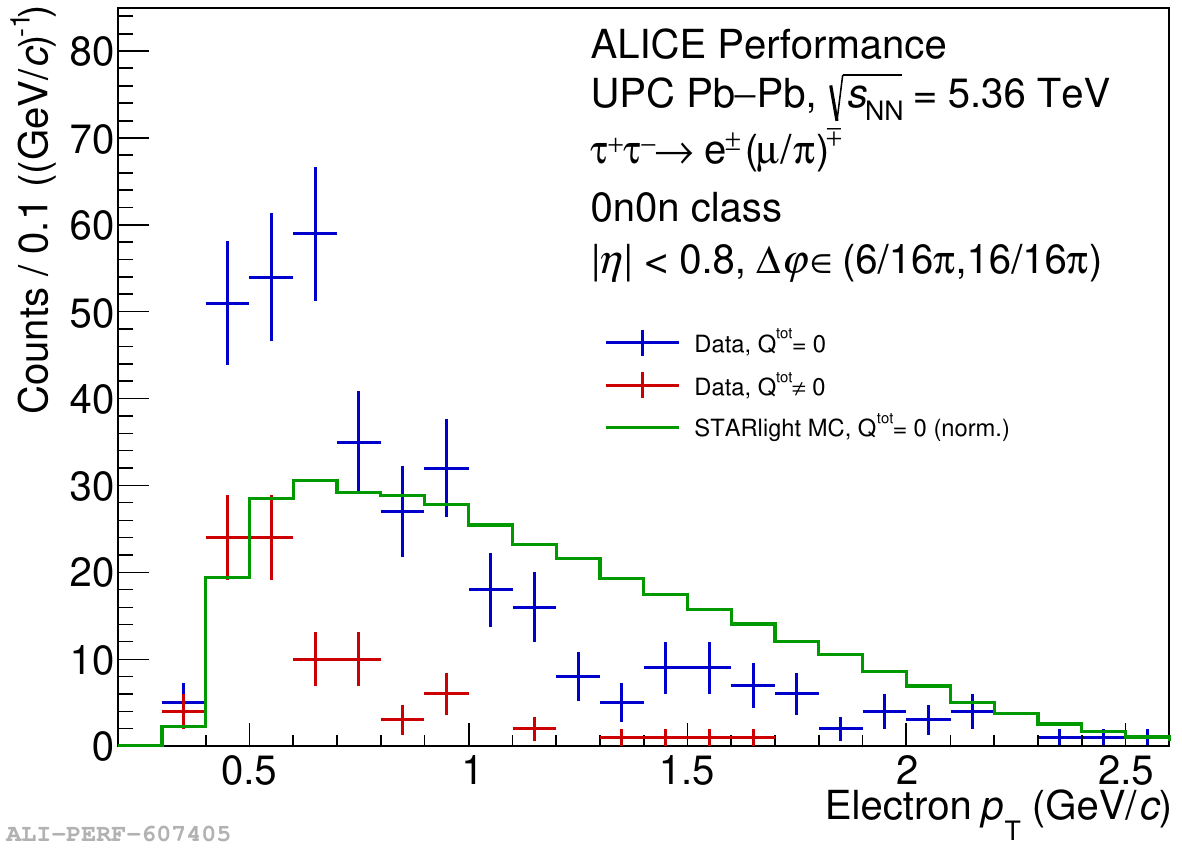}
  \caption{The electron \pT\ spectra in the \elel and $e+(\pi/\mu)$ selection compared to STARlight.}
  \label{fig:pt}
\end{figure}

Using the current \PbPb\ UPC dataset of $\sim\!1$~\nbinv\ at \snn$=\SI{5.36}{TeV}$, we report the following performance and observations:
\begin{itemize}
  \item \textbf{PID:} TPC $\mathrm{d}E/\mathrm{d}x$ shows clean electron separation with no indications of kaon/proton contamination in the selected samples; $\pi/\mu$ remain indistinguishable in the central barrel, consistent with expectations, see Figure~\ref{fig:pid}.
  \item \textbf{Acoplanarity:} Data agree with STARlight~\cite{Klein:2016yzr,Baltz:2009jk} MC in both $e^+e^-$ and $e+(\pi/\mu)$ selections after imposing a system-\pT\ requirement of $\SI{0.2}{\GeVc}$, see Figure~\ref{fig:acoplanarity}.
  \item \textbf{$e+3\pi$ channel:} A clear enhancement is observed for $m_{3\pi}<m_\tau$. Background-subtracted acoplanarity distributions reveal a signal consistent with MC expectations, see Figure~\ref{fig:fourprong}.
  \item \textbf{Electron \pT\ spectra:} The shapes in the $e^+e^-$ and $e+(\pi/\mu)$ channels are in fair agreement with STARlight predictions. Observed statistics match expectations given the delivered luminosity and selection efficiencies, see Figure~\ref{fig:pt}.
\end{itemize}

\section{Discussion and Outlook}
These first Run~3 results validate the UPC-based strategy to access $a_\tau$ with \ALICE. The central-barrel focus on low-\pT\ leptons provides sensitivity in a kinematic regime complementary to \pp\ analyses by CMS and ATLAS. With the anticipated Run~3+4 integrated luminosity of order $10$--$13$~\nbinv, \ALICE\ is projected to set world-leading constraints in the intermediate $W_{\gaga}$ range.

\section{Conclusions}
\begin{itemize}
  \item UPCs at the LHC enable a direct probe of the $\gamma\tau\tau$ vertex and the tau anomalous magnetic moment.
  \item \ALICE\ is uniquely capable of reconstructing low-\pT\ leptons in mixed tau-decay final states.
  \item First Run~3 data confirm detector performance and the analysis strategy, with distributions consistent with STARlight expectations.
  \item With the full Run~3+4 dataset, \ALICE\ can improve the current bounds on $a_\tau$ and offer coverage complementary to CMS/ATLAS in \pp.
\end{itemize}

\clearpage


\clearpage

\section*{Acknowledgments}

This work is supported by the Austrian Science Fund (FWF), Project No.~I~5277-N.

\end{document}